\documentclass[apj,twocolumn]{emulateapj}
\usepackage{natbib}
\usepackage{float}
\usepackage{amsmath}
\usepackage{epsfig}

\usepackage{graphicx}
\usepackage{subfigure}
\usepackage{color}
\definecolor{Blue}{rgb}{0.1,0.1,1.0}
\definecolor{Magenta}{rgb}{1.0,0.1,0.5}
\definecolor{LRed}{rgb}{0.8,0.0,0.0}

\newcommand{\nc}{\newcommand}

\nc{\be}[1]{\begin{equation}\mbox{$\label{#1}$}}
\nc{\bea}[1]{\begin{eqnarray} \mbox{$\label{#1}$}}
\nc{\Section}[2]{\section{#2}\label{#1}}
\nc{\Bibitem}[1]{\bibitem{#1}}
\nc{\Label}[1]{\label{#1}}

\nc{\eea}{\end{eqnarray}}
\nc{\ee}{\end{equation}}

\nc{\bdm}{\begin{displaymath}}
\nc{\edm}{\end{displaymath}}
\nc{\dpsty}{\displaystyle}
\nc{\bc}{\begin{center}}
\nc{\ec}{\end{center}}
\nc{\ea}{\end{array}}
\nc{\bab}{\begin{abstract}}
\nc{\eab}{\end{abstract}}
\nc{\btab}{\begin{tabular}}
\nc{\etab}{\end{tabular}}
\nc{\bit}{\begin{itemize}}
\nc{\eit}{\end{itemize}}
\nc{\ben}{\begin{enumerate}}
\nc{\een}{\end{enumerate}}
\nc{\bfig}{\begin{figure}}
\nc{\efig}{\end{figure}}

\nc{\arreq}{&\!=\!&}
\nc{\arrmi}{&\!-\!&}
\nc{\arrpl}{&\!+\!&}
\nc{\arrap}{&\!\!\!\approx\!\!\!&}
\nc{\non}{\nonumber}

\def\lsim{\; \raise0.3ex\hbox{$<$\kern-0.75em
      \raise-1.1ex\hbox{$\sim$}}\; }
\def\gsim{\; \raise0.3ex\hbox{$>$\kern-0.75em
      \raise-1.1ex\hbox{$\sim$}}\; }

\nc{\DOT}{\hspace{-0.08in}{\bf .}\hspace{0.1in}}
\nc{\Laada}{\hbox {$\sqcap$ \kern -1em $\sqcup$}}
\nc\loota{{\scriptstyle\sqcap\kern-0.55em\hbox{$\scriptstyle\sqcup$}}}
\nc\Loota{{\sqcap\kern-0.65em\hbox{$\sqcup$}}}
\nc\laada{\Loota}
\nc{\qed}{\hskip 3em \hbox{\BOX} \vskip 2ex}

\nc{\real}{{\rm I \! R}}
\nc{\Z}{{\sf Z \!\!\! Z}}
\nc{\complex}{{\rm C\!\!\! {\sf I}\,\,}}
\def\bigid{\leavevmode\hbox{\small1\kern-3.8pt\normalsize1}}
\def\id{\leavevmode\hbox{\small1\kern-3.3pt\normalsize1}}
\nc{\slask}{\!\!\!/}
\nc{\bis}{{\prime\prime}}
\nc{\pa}{\partial}
\nc{\ra}{\rangle}
\nc{\goto}{\rightarrow}
\nc{\swap}{\leftrightarrow}

\nc{\EE}[1]{ \mbox{$\cdot10^{#1}$} }
\nc{\abs}[1]{\left|#1\right|}
\nc{\at}[2]{\left.#1\right|_{#2}}
\nc{\norm}[1]{\|#1\|}
\nc{\abscut}[2]{\Abs{#1}_{\scriptscriptstyle#2}}
\nc{\vek}[1]{{\rm\bf #1}}
\nc{\integral}[2]{\int\limits_{#1}^{#2}}
\nc{\inv}[1]{\frac{1}{#1}}
\nc{\dd}[2]{{{\partial #1}\over{\partial #2}}}
\nc{\ddd}[2]{{{{\partial}^2 #1}\over{\partial {#2}^2}}}
\nc{\dddd}[3]{{{{\partial}^2 #1}\over
    {\partial #2 \partial #3}}}
\nc{\dder}[2]{{{d #1}\over{d #2}}}
\nc{\ddder}[2]{{{d^2 #1}\over{d {#2}^2}}}
\nc{\dddder}[3]{{d^2 #1}\over
    {d #2 d #3}}
\nc{\dx}[1]{d\,^{#1}x}
\nc{\dy}[1]{d\,^{#1}y}
\nc{\dz}[1]{d\,^{#1}z}
\nc{\dl}[1]{\frac{d\,^{#1}l}{(2\pi)^{#1}}}
\nc{\dk}[1]{\frac{d\,^{#1}k}{(2\pi)^{#1}}}
\nc{\dq}[1]{\frac{d\,^{#1}q}{(2\pi)^{#1}}}

\nc{\bfT}{{\bf T }}

\nc{\cA}{{\cal A}}
\nc{\cB}{{\cal B}}
\nc{\cD}{{\cal D}}
\nc{\cE}{{\cal E}}
\nc{\cG}{{\cal G}}
\nc{\cH}{{\cal H}}
\nc{\cL}{{\cal L}}
\nc{\cO}{{\cal O}}
\nc{\cT}{{\cal T}}
\nc{\cN}{{\cal N}}
\nc{\cR}{{\cal R}}

\nc{\rvac}[1]{|{\cal O}#1\rangle}
\nc{\lvac}[1]{\langle{\cal O}#1|}
\nc{\rvacb}[1]{|{\cal O}_\beta #1\rangle}
\nc{\lvacb}[1]{\langle{\cal O}_\beta #1 |}
\nc{\bb}{\bar{\beta}}
\nc{\bt}{\tilde{\beta}}
\nc{\ctH}{\tilde{\cal H}}
\nc{\chH}{\hat{\cal H}}
\nc{\al}{\alpha}
\nc{\g}{\gamma}
\nc{\Del}{\Delta}
\nc{\e}{\textrm{e}}
\nc{\eps}{\epsilon}
\nc{\lam}{\lambda}
\nc{\Om}{\Omega}
\nc{\ve}{\varepsilon}
\nc{\mn}{{\mu\nu}}
\nc{\vp}{\varphi}

\nc{\mL}{\mathcal{L}}
\nc{\rf}[1]{(\ref{#1})}
\nc{\nn}{\nonumber \\*}
\nc{\bfB}{\bf{B}}
\nc{\bfW}{\bf{W}}
\nc{\bfS}{\bf{S}}
\nc{\bfN}{\bf{N}}
\nc{\bfM}{\bf{M}}
\nc{\bfL}{\bf{L}}
\nc{\bfC}{\bf{C}}
\nc{\bfv}{\bf{v}}
\nc{\bfx}{\bf{x}}
\nc{\bfy}{\bf{y}}
\nc{\bfd}{\bf{d}}
\nc{\vx}{\vec{x}}
\nc{\vy}{\vec{y}}
\nc{\oB}{\overline{B}}
\nc{\oI}{\overline{I}}
\nc{\oR}{\overline{R}}
\nc{\rar}{\rightarrow}
\nc{\ti}{\times}
\nc{\slsh}{\hskip-5pt/}
\nc{\sm}{Standard~Model~}
\nc{\MP}{M_{\rm Pl}}
\nc{\mpl}{M_{\rm Pl}}
\nc{\tp}{t_{\rm Pl}}

\nc{\pmin}{p_{\rm min}}
\nc{\pmax}{p_{\rm max}}
\nc{\fo}{f_0}
\nc{\foi}{f_{0,i}\,}
\nc{\fop}{f_0^P}
\nc{\fou}{f_0^U}

\nc{\eff}{{\rm eff}}
\nc{\MT}{M_{\rm T}}
\nc{\ML}{M_{\rm L}}
\nc{\kk}{\vek{k}}
\nc{\pp}{{\rm p}}
\nc{\pt}{\partial_t}
\nc{\half}{{1\over 2}}
\nc{\w}{\omega}
\nc{\uhat}{\hat{U}_\w}
\nc{\hbn}{\hat{\bf{n}}}
\nc{\beq}{\begin{equation}}
\nc{\eeq}{\end{equation}}

\nc{\etal}{\mbox{\it et al.}}
\nc{\ie}{{\it i.e. }}
\nc{\eg}{{\it e.g. }}
\nc{\trh}{T_{\rm RH}}
\nc{\ad}{{a'\over a}}
\nc{\bd}{{b'\over b}}
\nc{\Rd}{{R'\over R}}
\nc{\diag}{{\textrm{diag}}}
\nc{\mato}[1]{\tilde{#1}}
\nc{\sinn}{\textrm{sinn}}
\nc{\sech}{\textrm{sech}}
\nc{\I}{\textrm{I}}
\nc{\II}{\textrm{II}}
\nc{\III}{\textrm{III}}
\nc{\vev}[1]{\langle #1 \rangle}
\nc{\hyp}{\,\; F_{1{\hskip -16pt}2}{\hskip 11pt}}
\nc{\brhom}{\overline{\rho}_M}
\nc{\brho}{\overline{\rho}}
\nc{\rhob}{\overline{\rho}}
\nc{\Pb}{\overline{P}}
\nc{\bH}{\overline{H}}
\nc{\ep}{{1+4\eps}}

\nc{\deriv}[2]{
\frac{\mathrm{d}#1}{\mathrm{d}#2}
}
\nc{\Mnu}{M_\nu}
\nc{\bee}{\begin{equation}}
\nc{\ene}{\end{equation}}
\nc{\hdp}{\sigma_8 (\Omega_{\rm m}/0.3)^{0.37}}
\nc{\avis}{\alpha_{vis}}
\nc{\cvis}{c^2_{vis}}
\nc{\clam}{c^2_{lam}}

\def\smiley{\hbox{\large$\bigcirc$\hspace{-.80em}%
\raise.2ex\hbox{$\cdot\cdot$}\kern-.61em    %--- .56
\lower.2ex\hbox{\scriptsize$\smile$}}\ }

\def\frowney{\hbox{\large$\bigcirc$\hspace{-.80em}%
\raise.2ex\hbox{$\cdot\cdot$}\kern-.635em
\lower.2ex\hbox{\scriptsize$\frown$}}\ }

\usepackage{graphicx}

\begin{document}

%\title{CMB Anomalies from Imperfect Dark Energy}
%\title{POSSIBLE TITLE: Confronting Anisotropic Dark Energy Models with Data}
\title{CMB Anomalies from Imperfect Dark Energy: Confrontation with the Data}
\author{Magnus Axelsson$^{1}$,  Frode Hansen$^{1}$, Tomi Koivisto$^{1,2}$, David
  F. Mota$^{1}$}
\affiliation{$^{1}$  Institute of Theoretical Astrophysics, University of
  Oslo, P.O.\ Box 1029 Blindern, N-0315 Oslo, Norway
  \\ $^{2}$ Institute for Theoretical Physics and the Spinoza Institute, Utrecht University,
Leuvenlaan 4, Postbus 80.195, 3508 TD Utrecht, The Netherlands }

\date{\today}

\begin{abstract}

We test anisotropic dark energy models with the 7-year WMAP temperature observations data.
In the presence of imperfect sources, due to large-scale gradients or anisotropies
in the dark energy field, the CMB sky will be distorted anisotropically on its way to us by the ISW effect.
%If dark energy is not a perfect fluid but for instance a vector field,
The signal covariance matrix then becomes nondiagonal for
small multipoles, but at $\ell \gtrsim 20$ the anisotropy is negligible. We parametrize possible violations of rotational invariance
in the late universe by the magnitude of a post-Friedmannian deviation from isotropy and its scale dependence.
This allows to obtain hints on possible imperfect nature of dark energy and the large-angle anomalous features in the CMB.
A robust statistical analysis, subjected to various tests and consistency checks, is performed to compare the predicted correlations
with those obtained from the satellite-measured CMB full sky maps. 
 The preferred axis point towards $(l,b) = (168^\circ, -31^\circ)$ 
and the amplitude of the anisotropy is $\varpi_0 = (0.51\pm 0.94)$ (1$\sigma$ deviation quoted). The best-fit model 
has a steep blue anisotropic spectrum ($n_{\mathrm{de}} = 3.1\pm1.5$).
\end{abstract}

\keywords{cosmic microwave background --- cosmology: observations ---
methods: numerical}

\maketitle

\section{Introduction}

In the past two decades great advances have been made in observational
cosmology. The most striking single discovery is the present acceleration of
the universe expansion, now confirmed by many independent experiments. The
most powerful probe of precision cosmology is the observations of the cosmic
microwave background (\cite{Bennett:2003bz,Hinshaw:2006ia}), which seem to
support the model of universe which at large scales is flat, isotropic and
homogeneous, as firmly predicted by inflation.  However, at more subtle level
there seems to be also hints of substantial anisotropy. Such would imply
violation of the cosmological principle, perhaps as striking change of paradigm
as the introduction of dark energy. Presently the evidence for anisotropy is
debatable, but the bounds can definitely be expected to improve with the
Planck experiment. Therefore it is extremely interesting to study theoretical
links between the acceleration and anisotropies, in particular, the
possibility to constrain them observationally \cite{Copi:2010na}.

Several distinct statically anisotropic features have been reported in the
data analysis of the CMB sky.  Among the most curious is the presence of
hemispherical asymmetry (\cite{Eriksen:2003db}). Recent investigations
exploiting the five-year WMAP data have found that the evidence for this
asymmetry is increasing and extends to much smaller angular scales than
previously believed to (\cite{Hansen:2008ym,Hoftuft:2009rq}). Alignment of the
quadrupole and octupole, the so called Axis of Evil (\cite{Land:2005ad}) could
also seem an unlikely result of statistically isotropic perturbations, even
without taking into account that these multipoles happen also to be aligned to
some extent with the dipole and with the equinox. In the CMB spectrum, the
angular correlation spectrum seems to be lacking power at the largest scales.
%This is the well known quadropole anomaly, which need not have a statistically anisotropic origin.  
The alignments seem to be statistically independent of the 
the lack of angular power (\cite{Rakic:2007ve}).  For 
other studies, see (\cite{Prunet:2004zy,Gordon:2005ai}).

It is natural to associate the apparent statistical anisotropy with dark
energy, since the anomalies occur at the largest scales, and these enter
inside the horizon at the same epoch that the dark energy dominance begins.
The paramount characteristic of dark energy is its negative pressure. One may
then contemplate whether this pressure might vary with the direction. Then
also the universal acceleration becomes anisotropic, and one would indeed see
otherwise unexpected effects.
%\footnote{Even in standard cosmology, one may expect such signatures due to the formation of local nonlinear structures. It seems though that to match the observations \cite{Rakic:2006tp}, a more complete understanding of the possibly more radical departures from standard cosmology taking into account the backreaction of nonlinearities is required; for recent progress see e.g. \cite{Rasanen:2009mg}.}  
These would presumably be strongest at the
smallest multipoles of the CMB since they describe the large angular scales
which are most directly affected during the late epochs of the
universe. Specifically, as the photons travel from the last scattering surface
towards us, the their temperature gets blue- and redshifted as they fall in
and climb out of the gravitational wells, respectively. When the potentials
evolve, there is a net effect in the temperature of the photons: this is the
integrated Sachs-Wolfe effect (ISW). Furthermore, if the average evolution of
the potentials was not the same in different directions of the sky, the effect
would be anisotropic. However, to explain the lack of large-angle correlations, 
there should occur a cancellation with the Sachs-Wolfe effect from the potentials 
last scattering surface that typically contribute to the large angles with similar order of magnitude
as the ISW \cite{Afshordi:2008rd}.

The potentials parameterising the perturbations of the metric, can be written in the longitudinal gauge as
\bee \label{long}
ds^2 = a^2(\eta)\left[-(1+2\psi)d\eta^2 + (1-2\phi)dx^idx_i\right].
\ene
The Poisson equation relates the spacetime curvature $\phi$ to the matter sources. As is well known, in the absence of anisotropic stress the
potentials $\phi$ and $\psi$ are equal. Thus, detection of inequality of these potentials in the present universe would indicate the
presence of imperfect energy source, either in form of dark energy fluid or modification of gravity (\cite{Koivisto:2005mm,Daniel:2008et,manera,dvali}).
Clearly, the difference of the potentials can be constrained much tighter at Solar system than at cosmological scales
(\cite{Mota:2007sz,Daniel:2009kr,ferreira,skordis,hu}). In the present study, we consider the possibility that the relation of the anisotropy described by the
difference of the potentials does not cancel out on the average, i.e. that the anisotropy is statistical. 

This amounts to promoting each Fourier mode of the potentials to depend not only on the length but also on the direction
of the wavevector. This is a generic prediction for perturbations in a non-FRW universe and also for non-scalar field models, in particular vector
fields \cite{ArmendarizPicon:2004pm,Koivisto:2008ig,vec1,Li}.
It has been considered if the cosmological fine-tunings could be more naturally alleviated with a dynamical dark 
energy component when this is modelled with a more general field than a scalar. Vector fields dynamics 
could accelerate
the universe today (\cite{Kiselev:2004py,Jimenez:2008au}) having phantom evolution without UV pathology (\cite{Rubakov:2006pn,Libanov:2007mq})
possibly connecting the acceleration with the electromagnetic scale (\cite{Jimenez:2008nm,Jimenez:2009dt,vec10}).

There has also been interest on anisotropies in inflation (\cite{Gumrukcuoglu:2006xj,Ackerman:2007nb,boehmer}), and their 
comparison with the data (\cite{Groeneboom:2008fz,Groeneboom:2010hp, ArmendarizPicon:2008yr}). The anisotropy in the primordial spectrum could be generated by vector fields 
(\cite{Golovnev:2008cf,Koivisto:2008xf}) or more general
n-forms (\cite{Germani:2009iq,Koivisto:2009sd,Koivisto:2009fb}). In particular this can be robustly realized by the vector curvaton paradigm 
(\cite{Dimopoulos:2009am}) which \cite{Dimopoulos:2011pe} recently implemented within D-brane inflation in 
type II string theory by taking into account the $U(1)$ gauge field that lives on the brane. 
Perturbations have been also studied in anisotropically 
inflating backgrounds (\cite{Pereira:2007yy,Gumrukcuoglu:2007bx}) and in the shear-free cosmologies considered in \cite{zspace,Zlosnik:2011iu}, where 
the expansion is isotropic but the spatial curvature depends on the direction. 
These homogeneous but anisotropic universes could emerge by tunneling from a lower-dimensional vacuum \cite{Adamek:2010sg,Graham:2010hh}. 

%This is qualitatively different from shear and vorticity, which correspond to anisotropy of the expansion or the rotation of the universe.

In the presence of such variety of possibilities, we choose to rather employ a general parametrisation than 
study a particular model. To that purpose, we parameterize directly the angular variation of the gravitational 
potentials. This can be seen as a step towards
a more complete anisotropic post-Friedmannian parametrisation of the deviations from standard GR $\Lambda$CDM cosmology, inspired by the recent development of a fully consistent parametrisation encompassing statistically isotropic models (\cite{ferreira,Baker:2011jy}).
In Section \ref{sec:cmb} we derive the signal covariance matrix in the presence of generalised perturbation sources.
In Section \ref{sec:ade} we describe our parametrisation of such sources and their interpretation as a as anisotropies of the dark energy field or as some
spontaneous anisotropisation of the CMB radiation.
In Section \ref{sec:analysis} we discuss the analysis, and the method we use is described in detail in Section \ref{sec:analysis}, 
and finally, the results are presented in Section \ref{sec:results}.

\section{CMB from anisotropic scalar sources}
\label{sec:cmb}

The temperature anisotropy field is conventionally expanded in terms of the spherical harmonics and on the other
hand considered in the Fourier space
\bee \label{exps}
\Theta({\bf x}, \hat{e},\eta) = \sum_{\ell=0}^{\infty}\sum_{m=-\ell}^{\ell}a_{\ell m} Y_{\ell m} =
\int \frac{d^3k}{(2\pi)^3} e^{i{\bf k \cdot x}} \delta({\bf k}) \Theta({\bf k},\hat{e}, \eta),
\ene
where we have normalized the transfer function $\Theta({\bf k},{\bf e},\eta)$ with respect to
the initial amplitude $\delta({\bf k})$. One makes contact between the two expansions by using the Rayleigh formula
\bee \label{rayleigh}
e^{i{\bf x}\cdot {\bf k}} = \sum_{\ell=0}^{\infty} i^\ell (2\ell+1) j_\ell(kx) Y^*_{\ell m}(\hat{k})Y_{\ell m}(\hat{x}),
\ene
together with the addition theorem for the spherical harmonics
\bee \label{addition}
P_\ell(\hat{k}\cdot\hat{p}) = \frac{4\pi}{2\ell+1}\sum_{m=-\ell}^\ell Y^*_{\ell m}(\hat{k}) Y_{\ell' m'}(\hat{p})
\ene
for the second equality in Eq.(\ref{exps}) and then, by exploiting the orthonormality of the spherical harmonics,
picks up the coefficients in the first equality in Eq.(\ref{exps}). These are
\bee \label{alm}
a_{\ell m} = i^\ell \int \frac{d^3k}{2\pi^2}\delta({\bf k}) Y^*_{\ell m}(\hat{k})\Theta_l({\bf k}).
\ene
where we have defined
%\footnote{The definitions of the Legendre expansion coefficients $\Theta_\ell$ sometimes differ by
%an absorbtion of $i^\ell$. Multiplying this would lead to inconsistency of the results (the cross-term correlators below
%would become antisymmetric and the pure anisotropy correlators would differ by a sign from the previous results when $\ell' = \ell \pm 2$).}
\bee
\Theta_l({\bf k}) = \int j_\ell (kr(\eta))\Theta({\bf k},\eta)d\eta.
\ene
We assume, as usual, that the primordial spectrum of perturbations is statistically isotropic,
\bee
\langle \delta({\bf k})\delta^*({\bf k'}) \rangle = P(k)(2\pi)^3\delta^3({\bf k}-{\bf k'}).
\ene
However, we allow the transfer function an anisotropic part,
\bee \label{transfer}
\Theta_l ({\bf k}) = \Theta^0_\ell(k) + \omega(\hat{k}\cdot \hat{n})\Theta^A_\ell(k).
\ene
The second term can then incorporate the anisotropic ISW contribution from dark energy.
We are then interested in the correlators
\bee \label{correlators}
\langle a_{\ell m} a^*_{\ell' m'}\rangle = \frac{2i^{\ell-\ell'}}{\pi}\int d^3k P(k) Y^*_{\ell m}(\hat{k}) Y_{\ell' m'}(\hat{k})
\Theta_l ({\bf k})\Theta_{l'}^* ({\bf k}).
\ene
The expression follows directly from Eq.(\ref{alm}). One may arrive at the same result by inverting Eq.(\ref{exps}) to obtain
the $a_{\ell m}$, expanding $\Theta({\bf k},\hat{e}, \eta)$ as a Legendre series and using the addition theorem (\ref{addition})
to eliminate the Legendre polynomials $P_\ell$ when integrating over the direction in the sky.

It is useful introduce the spherical components as in \cite{Ackerman:2007nb} of the direction vector
\bee \label{vector}
n_\pm = \mp \left(\frac{\hat{n}_x \mp i \hat{n}_y}{\sqrt{2}}\right), \quad n_0 = \hat{n}_z,
%n_\pm = \frac{1}{\sqrt{2}} \left(\mp \hat{n}_x + i \hat{n}_y\right), \quad n_0 = \hat{n}_z,
\ene
since then one may write
\bee
\hat{k}\cdot \hat{n} = 2\sqrt{\frac{\pi}{3}}\left[n_+ Y^{+1}_1(\hat{k}) + n_- Y^{-1}_1(\hat{k}) + n_0 Y^{0}_1(\hat{k})\right].
\ene
We arrive at
\bee \label{corr}
\langle a_{\ell m} a^*_{\ell' m'}\rangle = \frac{2i^{\ell-\ell'}}{\pi}\left[\delta_{m',m}\delta_{\ell',\ell}I_\ell
+ \zeta_{\ell m;\ell' m'}I^A_{\ell\ell'} +  \xi_{\ell m;\ell' m'}I^{AA}_{\ell\ell'}\right].
\ene
Here the source integrals are
\bee \label{i0}
I_\ell = \int_0^\infty dk k^2 P(k)  \left[ \Theta^0_\ell(k) \right]^2,
\ene
\bee \label{ia}
I^A_{\ell\ell'} = \int_0^\infty dk k^2 P(k)  \left[\omega\Theta^0_{\ell'}(k) \Theta^A_\ell(k)
                                               + \omega^*\Theta^0_\ell(k) \Theta^A_{\ell'}(k)  \right],
\ene
and
\bee \label{iaa}
I^{AA}_{\ell\ell'} = \int_0^\infty dk k^2 P(k) |\omega|^2 \Theta^A_\ell(k) \Theta^A_{\ell'}(k).
\ene
The first term in Eq.(\ref{corr}) is the isotropic contribution. The second is the
cross term, for which the geometric coefficients are given by
\bee
\zeta_{\ell m;\ell' m'} = n_+ \zeta^+_{\ell m;\ell' m'} + n_-\zeta^-_{\ell m;\ell' m'} + n_0\zeta^0_{\ell m;\ell' m'},
\ene
where
\begin{widetext}
\bee \label{zeta+}
\zeta^+_{\ell m;\ell' m'} = \delta_{m',m-1}\left[
\delta_{\ell',\ell-1} \sqrt{\frac{{(\ell+m-1)(\ell+m)}}{2(2\ell-1)(2\ell+1)}} -
\delta_{\ell',\ell+1} \sqrt{\frac{{(\ell-m+1)(\ell-m+2)}}{2(2\ell+1)(2\ell+3)}} \right],
\ene
\bee \label{zeta-}
\zeta^-_{\ell m;\ell' m'} = \delta_{m',m+1}\left[
\delta_{\ell',\ell-1} \sqrt{\frac{{(\ell-m-1)(\ell-m)}}{2(2\ell-1)(2\ell+1)}} -
\delta_{\ell',\ell+1} \sqrt{\frac{{(\ell+m+1)(\ell+m+2)}}{2(2\ell+1)(2\ell+3)}} \right],
\ene
\bee \label{zeta0}
\zeta^0_{\ell m;\ell' m'} = \delta_{m',m}\left[
\delta_{\ell',\ell-1} \sqrt{\frac{{(\ell-m)(\ell+m)}}{(2\ell-1)(2\ell+1)}} +
\delta_{\ell',\ell+1} \sqrt{\frac{{(\ell-m+1)(\ell+m+1)}}{(2\ell+1)(2\ell+3)}} \right].
\ene
\end{widetext}

One can check that $\zeta^*_{\ell m;\ell' m'} = \zeta_{\ell' m';\ell m}$.
The last term in Eq.(\ref{corr}) is the autocorrelation of the anisotropic piece.
The geometric coefficients $\xi_{\ell m;\ell' m'}$ have been previously presented in \cite{Ackerman:2007nb}. We have them with an extra
minus sign for the off-diagonal components\footnote{The reason for this discrepancy was a forgotten $i^{\ell-\ell'}$ factor in the Ackerman paper.}.
The factor $i^{\ell-\ell'}$ also results in odd-parity correlations being imaginary. 

Although our cosmology features anisotropies, we assume that the underlying model is Gaussian. For a pedagogic discussion of these statistical properties and their
tests, see \cite{Abramo:2010gk}.

\section{Anisotropically stressed dark energy}
\label{sec:ade}

The main reason for disregarding the anisotropic stress in the dark energy
fluid might be that a minimally coupled scalar field, conventional parametrisation of the inflationary energy source, cannot generate anisotropic stresses.
However, since there is no fundamental theoretical model to describe dark energy, one might miss the behind physics of acceleration by sticking to the assumption
of zero anisotropic stress. Such stresses are quite by viscous fluids, any higher spin fields and non-minimally coupled scalar fields too \cite{bat1,bat2,ani1,ani2,ani3,ani4,ani5,ani6}. To study such a generic
property with a many possible theoretical realizations, it is useful to employ a parametrisation of its physical consequences.

An efficient way to describe possible deviations from perfect-fluid cosmology is to introduce
the post-general relativity cosmological parameter $\varpi$ along the lines of \cite{Caldwell:2007cw},
which is defined as the difference of gravitational potentials in the Newtonian gauge,
\bee \label{psi}
\psi = (1+\varpi)\phi,
\ene
where the line element reads
\bee
ds^2 = a^2(\eta)\left[-(1+2\psi)d\eta^2 + (1-2\phi)dx^idx_i\right],
\ene
This parameter then appears as a cosmological generalization of the post-Newtonian $\gamma$,
for which one has tight constraints from the Solar system scales \cite{Will:2001mx}. An economic assumption is
then that such a generalized parameter depends, at all relevant scales, on the ratio of matter and
dark energy densities,
\bee
\varpi = \varpi_0 \frac{\rho_{DE}}{\rho_M}.
\ene
Then one has reasonable constraints on the $\varpi_0$ from various scales, and $\varpi_0$ of order
one would imply a variety of phenomenology at different scales, ranging from Solar system physics
to cosmology, just at the verge of detection. Here we study on cosmological effects of dark energy
and adopt the recipe 3 of \cite{Caldwell:2007cw} as the basis to parametrize the
shear stress of dark energy.

In particular, we will explore the case that the anisotropic stress has a preferred direction as in \cite{Koivisto:2008ig}.
For each Fourier mode of cosmological perturbations, we write
\bee \label{varpi}
\varpi = i(\hat{k}\cdot\hat{n})\varpi_0 \frac{\rho_{DE}}{\rho_M},
\ene
where $\hat{n}$ is the direction of the anisotropy and $\varpi_0$ is real. Consider then the transfer function Eq.(\ref{transfer}).
The $\Theta^0_\ell(k)$ would now be as usual. Thus it includes contributions from both early and late universe. At the largest scales the
Sachs-Wolfe effects are known to dominate the anisotropy sources. The second part would be given
by the rotationally non-invariant part of the ISW contribution, which in our present prescription is the following:
\bee \label{extra}
\Theta^A_\ell(k) = - i\varpi_0 \int e^{-\kappa (\eta)} \frac{d}{d\eta}\left(\frac{\rho_{DE}}{\rho_M}\phi_k(\eta)\right) j_l[kr(\eta)]d\eta,
\ene
where $\phi_k(\eta)$ is given by the standard computation. We just add the extra contributions due to Eq.(\ref{extra}) to the
sources from which to compute the correlators as described in the previous section.
Therefore it becomes straightforward to determine the features in the CMB sky in this prescription.

This parametrization describes a gradient-type modification of the effective CMB sources. We note that \cite{Tangen:2009mw} has determined the implications of a super-horizon
perturbation, and \cite{Erickcek:2008sm} considered such a spatial variation of the curvaton field at inflation.
In our model, the anisotropy is formed dynamically and becomes important with the dominance of dark energy.
Explicitly, we have a spontaneous modification of the effective CMB sources through the gradient operator as follows:
\bee
\psi({\bf x}) = \left[1 + ({\bf n}\cdot \nabla)\right] \phi({\bf x})
\ene
when $|n| = \varpi_0 (\rho_{DE}/\rho_M)$. This amounts to shifting the Fourier modes of the perturbations exactly as prescribed in (\ref{psi}) and
(\ref{varpi}),
\bee
\psi_{\bf k} = \left[1 + i(\hat{k} \cdot \hat{n})\varpi_0\frac{\rho_{DE}}{\rho_D}\right]\phi_{\bf{k}}.
\ene
Since the anisotropic part develops as a result of the evolution of the universe, it is a property of the transfer functions and not of the primordial
spectrum of of perturbations. One does not expect odd $\Delta\ell$ couplings from primordial origin, since they violate parity.
The presence of the imaginary unit is necessary for reality of the perturbations,
which can be checked as follows. Since the physical perturbation is a convolution of the primordial physical perturbation
and the Fourier transformation of the transfer function, one notes that the latter should also be real. We get that
\begin{eqnarray}
\frac{\psi({\bf x})}{\psi_{Primordial}({\bf x})} &=&
\frac{1}{2\pi^3}\int_0^\infty dk \frac{k}{x}\sin{(kx)} \\ \nonumber &&\left[
  1+(\hat{x}\cdot\hat{n})\left(\frac{1}{kx}-\cot{(kx)}\right)\right]\phi(k),
\end{eqnarray}
where $\phi(k)$ is the (real, isotropic) transfer function which depends only on the magnitude of the wavevector, and the right hand
side is the anisotropic transfer function in the configuration space that retains it's reality.

To recap our approach, we introduced a mismatch of the two gravitational potentials in the Newtonian gauge. This mismatch, quantified by $\varpi$, is given by a gradient along a
preferred axis $\hat{n}$. In the following, we will also allow scale dependence of this effect by introducing the spectral index $n_{\mathrm{de}}$. This parametrisation can then be used to
constrain the presence of the such gradients in the late universe, since they would be seen in the CMB (practically only)
through their impact on the time-evolution of the gravitational potentials because of the ISW effect. Physically, these gradients could be caused by a large-scale inhomogeneity
entering our horizon, spontaneous formation due to e.g. coherent magnetic fields or simply the possible imperfect nature of the dark energy field.

To clarify the difference of our approach to all previous literature, let us mention that odd modulations may be considered to occur at three distinct levels. The temperature field itself can be modulated, for a recent example see \cite{Aluri:2011wv}. This would effectively describe some systematics in the data. Strangely enough, the primordial spectrum itself could contain parity violating contribution \cite{Koivisto:2010fk}. That is consistent only in the context of noncommutative quantum field theory, and thus provides a unique signal for such high energy modifications of the 
standard model \cite{Groeneboom:2010fn}. Finally, the cosmological structures may evolve statistically oddly, which is the case we focus upon here.

\section{Model fitting}
\label{sec:analysis}

In this section we describe in detail the confrontation of the model with the WMAP data. We will now discuss the method used to obtain the set of parameters which gives the best fit 
between our model and the observations. Our basis is the evaluation of the Likelihood function in a 4-dimensional parameter space. After explaining the general likelihood procedure, 
we explain in more detail the different steps taken to calculate and maximize the likelihood.

\subsection{Data model and notation}
\label{sec:dataandnotation}

Given a set of data $\{d_i\}$ our goal is to find the set of parameters which maximizes the posterior.
For ease of notation let $\alpha = (\theta,\phi,\varpi_0,n_{\mathrm{de}}) = (\hbn,\varpi_0,n_{\mathrm{de}})$ denote the set
of parameters to be determined. By Bayes' theorem we know that the posterior distribution $P(\alpha|\bfd) \propto P(\bfd |\alpha)
P(\alpha) = \mL(\alpha)P(\alpha)$ where $\mL(\alpha)$ is the likelihood and $P(\alpha)$ is a prior.
We take a conservative approach and assume that we know nothing prior about the anisotropic parameters, and thus 
$P(\alpha|\bfd) = \mL(\alpha)$ up to a normalization constant. If we manage to compute the likelihood function in
all of the parameter space then we automatically have the posterior distribution and our job is essentially done.

Although our model is anisotropic we still assume that the underlying distribution is Gaussian, and thus the likelihood is
\beq
\label{eq:likelihood}
\mL(\al) \propto \frac{e^{-\frac{1}{2}\bfd^\dagger \bfC^{-1}(\al)\bfd}}{\sqrt{\mathrm{det}\bfC(\al)}}
\eeq
where the data vector $\bfd$ consists of the $a_{\ell m}$ of the observed masked map. 
The correlation matrix $\bfC=\bfS+\bfN$ is the sum of the CMB signal covariance matrix $\bfS$ and the noise covariance matrix $\bfN$. Our analysis is performed
in harmonic space where the signal covariance $S_{\ell m;\ell' m'} = \langle a_{\ell m}a^*_{\ell' m'}\rangle$ is computed
from Eq. (\ref{corr}) and thus contains non-diagonal anisotropic contributions from dark energy. This matrix gives
the dependence on the cosmological parameters.

Finally the observed data vector $\bfd$ may be written in harmonics space as
\beq
\mathrm{d}_{\ell m} = \mathrm{b}_{\ell}\mathrm{w}_\ell s_{\ell m} + \mathrm{n}_{\ell m}
\eeq
where b$_\ell$ is the instrumental beam, w$_\ell$ is the pixel window function and n$_{\ell m}$ is the (Gaussian) noise term. 
Since there is no correlation between the signal and the noise we have
\beq
\langle d_{\ell m}d^{*}_{\ell' m'} \rangle = \langle \tilde{s}_{\ell m}\tilde{s}^{*}_{\ell' m'} \rangle + \langle \mathrm{n}_{\ell m}\mathrm{n}^{*}_{\ell' m'} \rangle
\eeq
where $\tilde{s}_{\ell m} = \mathrm{b}_{\ell}\mathrm{w}_\ell s_{\ell m}$ is the observed signal. The goal is now to maximize this likelihood with respect to the model parameters, 
but we will first explain in some detail how we calculate the covariance matrices involved in the likelihood calculation.

\subsection{Signal covariance}

From equations (\ref{corr}-\ref{zeta0})
we see that the covariance matrix, in addition to the diagonal isotropic contribution $I_{\ell}$ contains
the cross term coefficients $\zeta_{\ell m; \ell' m'}$ which couple $\ell$ to $\ell' = [\ell \pm 1]$ and $m$ to
$m' = [m, m \pm 1]$. The last term $\xi_{\ell m; \ell' m'}$ is the Ackerman \citep{Ackerman:2007nb} term for which
we have couplings when $\ell' = [\ell, \ell \pm 2]$ with $m' = [m, m \pm 1, m \pm 2]$. All other terms are
zero.

\begin{figure}[t]
  \mbox{\epsfig{figure=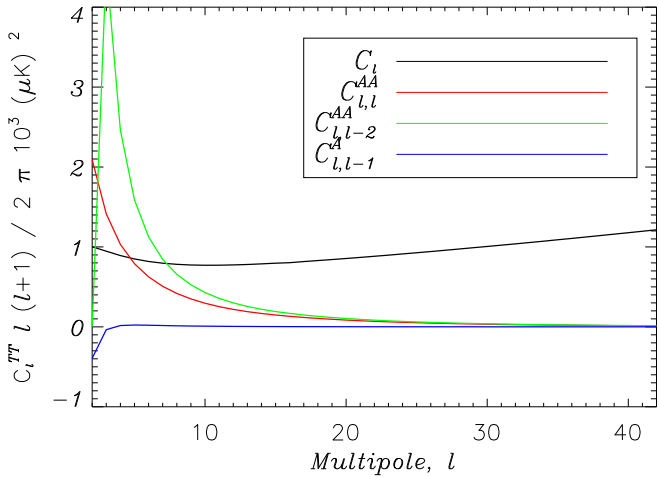,width=\linewidth,clip=}}
  \caption{(Unnormalized) Integrals (eq. \ref{cl0} - \ref{claa}) computed from a modified version of CAMB \cite{Lewis:1999bs}. Notice how the anisotropic integrals decay
towards zero after only a few multipoles. These integrals are all used in the construction of the signal
covariance matrix. An anisotropic scalar spectral index of $n_{\mathrm{de}} = 1.0$ is used in this plot, and we have not normalized them.
(Color version of this figure is available online)}
\label{fig:cls}
\end{figure}

We will now express the integrals in  equations (\ref{zeta+}-\ref{zeta0}) in terms of power spectra $C_\ell$, $C^{A}_{\ell\ell'}$, $C^{AA}_{\ell\ell'}$ as
\beq
\label{cl0}
C_\ell=\frac{2}{\pi}I_{\ell \ell}
\eeq
\beq
\label{cla}
C^A_{\ell\ell'}=\frac{2}{\pi}I^A_{\ell \ell'}
\eeq
\beq
\label{claa}
C^{AA}_{\ell\ell'}=\frac{2}{\pi}I^{AA}_{\ell \ell'}
\eeq
From equations (\ref{zeta+}-\ref{zeta0}) we see that $C^{A}_{\ell\ell'}$ and $C^{AA}_{\ell\ell'}$ only give contributions for $\ell'=\ell\pm1$ and $\ell'=\ell\pm2$. 
In figure \ref{fig:cls} we have calculated (using a modified version of CAMB) and plotted these integrals (because of symmetry the $\ell'=\ell+1$ and $\ell'=\ell+2$ 
terms are equal to the $\ell'=\ell-1$ and $\ell'=\ell-2$ terms plotted) and compared to the isotropic power spectrum. From this figure we clearly see that the
anisotropic contribution from the dark energy component becomes negligible except at the largest scales where
the anisotropic contribution even exceeds the isotropic. The most prominent altercation comes from the anisotropic integral $C^{AA}_{\ell,\ell-2}$ contributing to the
off-diagonal elements of the signal covariance matrix.
Due to the short range of the anisotropic integrals we have altered the pivot scale in CAMB from $k_0 = 0.05 \,\mathrm{Mpc}^{-1}$ to 
$k_0 = 2\times 10^{-3} \,\mathrm{Mpc}^{-1}$. In this way, we ensure that the spectral index enters correctly to tilt these integrals. 

\subsection{Transformation of variables}
\label{sect:trans}
We have noticed from simulations that there was a significant degeneration between the anisotropic spectral index $n_{\mathrm{de}}$ and the 
amplitude $\varpi_0$ due to the fact that they both regulate the magnitude of the non-diagonal signal matrix elements. 
In order to ease the estimation procedure, we choose to estimate for a new variable $\varpi^{a}_0$ instead of $\varpi_0$ defined by 
$\varpi^{a}_0 = \varpi_0/\sqrt{a}$ where we define $a$ as $a = A(I^{\mathrm{AA}}_{\ell \ell-2}(1.0))/A(I^{\mathrm{AA}}_{\ell \ell-2}(n_{\mathrm{de}}))$ where 
 $A(I^{\mathrm{X}}_{\ell \ell'} (n_{\mathrm{de}}))$ is the area under the anisotropic integral $I^{\mathrm{X}}_{\ell \ell'}(n_{\mathrm{de}})$ where $X=\{A,AA\}$ and $n_{\mathrm{de}}$ is the dark 
energy spectral index. The parameters  $n_{\mathrm{de}}$  and $\varpi^{a}_0$ are not degenerate and can thus more easily be estimated for. In the end, we convert to the physical
parameter $\varpi_0$ and all results are quoted in terms of this parameter.

\subsection{Modification of spectrum}
\label{sect:modspec}
The angular power spectrum $C_\ell$ will receive a contribution from the anisotropy which could have an
observable impact on the largest scales of the universe. This can be seen from equations (\ref{zeta+}-\ref{zeta0})
and the form of the $\xi_{\ell m ; \ell' m'}$ elements (see \cite{Ackerman:2007nb} for details). One must therefore be careful when performing the full
analysis, making sure that any choice for $\varpi_0$ does not significantly affect the power spectrum away from the WMAP best fit spectrum, but
only the anisotropic contribution to the correlations between $a_{\ell m}$s. 

To quantize these statements we calculate the net extra power from the anisotropic contribution. It is
given by the diagonal part of the anisotropy which can be written
\beq
\Delta C_{\ell} = \frac{2 I^{AA}_{\ell \ell}}{\pi}\frac{\varpi_0^2}{2\ell +1} \sum_{m=-\ell}^{\ell} \xi_{\ell m; \ell m}
\eeq
Using the explicit form of $\xi_{\ell m; \ell m}$ to perform the summation we find the modified power spectrum 
(see the Appendix for details):
\beq
\label{clmod}
C_{\ell}^{\mathrm{mod}} = C_{\ell} + \frac{2}{3\pi}\varpi_0^2 I^{\mathrm{AA}}_{\ell \ell}(n_{\mathrm{de}}) \equiv
C_\ell +\frac{\varpi^2_0}{3}C^{\mathrm{AA}}_\ell
\eeq
The extra contribution to the power-spectrum from the anisotropic sources depends on the amplitude parameter
$\varpi_0$ (degree of isotropy breaking) and the dark energy spectral index (parameterisation of the fluid scale-
dependence) which we have included explicitly as an argument in $I^{\mathrm{AA}}_{\ell \ell}$. 

We see from equation \ref{clmod} that there is a limit to what values the amplitude $\varpi_0$ may take in order
to obtain a power spectrum which is consistent with the WMAP7 best fit. This is however only true when we assume the
other cosmological parameters to have the WMAP best fit values. Clearly the new parameters which we have introduced
allows for the other parameters to vary and one should re-estimate the other cosmological parameters together with
$\varpi_0$ and $n_\mathrm{de}$. In order to test the anisotropic model without running a full cosmological parameter
estimation one may renormalize the covariance matrix for a given set of parameters ($\varpi_0$, $n_\mathrm{de}$) in 
such a way that the power spectrum is kept constant at the best fit WMAP model. In this case the signal covariance becomes
\beq
\label{normSignal}
S^{Norm}_{\ell m;\ell'm'} = \left(\frac{S_{\ell m;\ell'm'}}{\sqrt{S_{\ell m;\ell m}S_{\ell'm';\ell'm'}}}\right)C_\ell
\eeq
where $S_{\ell m;\ell m} = C_\ell + \xi_{\ell m;\ell m}\varpi^2_0C^{\mathrm{AA}}_\ell$.
With this normalization, the diagonal part of our signal covariance matrix will match the WMAP power spectrum
regardless of amplitude and spectral index for the dark energy, while the off-diagonal components describe relative
anisotropy.

\subsection{Noise covariance}
\label{sec:noise}
The noise in pixel space is assumed to be uncorrelated between pixels, \ie
$\mathbf{N}_{ij} = \langle n_i n_j \rangle = \sigma^2_i \,\delta_{ij}$ where $i$ and $j$ are pixel indices, and $\sigma_i$
is the noise root-mean-square deviation. The noise covariance matrix in pixel space is therefore diagonal. When going to spherical harmonic space, 
the harmonic coefficients of the noise are correlated and $N_{\ell m;\ell' m'} = \langle n_{\ell m} n^{*}_{\ell' m'} \rangle$ is therefore a dense matrix.

Expanding the noise harmonic coefficients in terms of pixel space quantities we
eventually find that the expression for the noise matrix in harmonic space becomes
\begin{widetext}
%\begin{align}
  \beq
  \label{eq:Nlmmat}
  N_{\ell_1 m_1 ; \ell_2 m_2} = (-1)^{m_{1}}\sqrt{\frac{(2\ell_1+1)(2\ell_2+1)}{4\pi}}%\nonumber \\
  \sum_{\ell_3=0}^{2\ell_{\mathrm{max}}}a_{\ell_3 m_3} \sqrt{2\ell_3+1}
  \left(\begin{array}{ccc}
      \ell_3 & \ell_1 & \ell_2 \\
      \\
      \,0\quad & \,\,0\quad & \,0\,
    \end{array}\right)
  \left(\begin{array}{ccc}
      \ell_3 & \ell_1 & \ell_2 \\
      \\
      m_3\quad    & -m_1\,    & m_2\,
    \end{array}\right)
  %\nonumber \\
  \delta(m_3 - m_1 + m_2)
%\end{align}
  \eeq
\end{widetext}
where the Dirac delta-function arises due to the requirement $m_1 - m_2 - m_3 = 0$ for the Wigner 3$j$ symbols. It is also required
that the triangle condition $|\ell_1 - \ell_2| \leq \ell_3 \leq \ell_1 + \ell_2$ is fulfilled. Notice also that due to this relation
the sum over $\ell_3$ goes up to $2\ell_{\mathrm{max}}$.

The $a_{\ell_3 m_3}$ coefficients originate from a spherical transform of the variance of the noise map, $\sigma_{i}^2$. 
Eq. (\ref{eq:Nlmmat}) is then implemented into our code. It needs only be computed once for each run of the code and
is added to the signal covariance matrix in the step before the skycut is applied.

\subsection{Correlations introduced by the mask}

If we let $C_{\ell m; \ell' m'}$ denote the covariance matrix without mask, and $\tilde{C}_{\ell m; \ell' m'}$ denote the corresponding matrix
including correlations from the sky cut, then the relation between them in harmonic space is found to be
\beq
\tilde{C}_{\ell m; \ell' m'} = \sum_{L M}\sum_{L'M'} W_{\ell m; LM} C_{LM;L'M'} W^{*}_{\ell' m';L'M'}
\label{eq:skyrel}
\eeq
which can be written compactly in matrix form as $\tilde{\bfC} = \bfW \bfC \bfW^{\dagger}$.
The operation in
Eq. (\ref{eq:skyrel}) can be shown to be additive so the covariance matrix is the sum of the signal plus noise correlation matrices.
The multipole range here is $L,L' \in [2, \ell_{\mathrm{max}}]$, and the sums over $M,M'$ here run over positive values. 
The hermitean coupling matrix $W_{\ell m; \ell' m'}$ defined by
\beq
\label{eq:cpK}
W_{\ell m; \ell' m'} = \int d\Omega \, M(\Omega)Y^{*}_{\ell m}(\Omega)Y_{\ell' m'}(\Omega)
\eeq
is a function of the pixel space mask $M(\Omega)$ (where $\Omega = (\theta,\phi)$ is the angular position on the sky) and so depends on the resolution 
$N_{\mathrm{side}}$. It quantifies the new couplings between modes that arise due to the fact that we are now not analyzing a full sky.

Starting with the WMAP KQ85 mask at $N_{\mathrm{side}}=512$ we degrade our mask so that the operation of applying the mask in pixel space can be traced exactly 
by applying the kernel matrix $\bfW$ in harmonic space. This is done by first smoothing with a Gaussian beam of fwhm$ = 744$ arcmin, and then setting 
$M(p) = 0$ (where $p$ is a HEALPix pixel index) where $M(p) < 0.80$. We then band-limit the mask so that it contains multipoles in the desired range 
\citep{ArmendarizPicon:2008yr}. 
These operations ensure that our mask does not contain small-scale structures. In the process the mask is expanded so that it now covers about $25\%$ 
of the sky.

It now remains to give an expression for the coupling kernel. Eq. (\ref{eq:cpK}) can be transformed by decomposing the mask into
spherical harmonics and then performing the resulting integral over all angles to obtain again Wigner $3j$ symbols. This is exactly the same 
analytical procedure which led to Eq. (\ref{eq:Nlmmat}) with some minor modifications. The result in this case becomes
\begin{align}
  \label{eq:Wlmmat}
  &W_{\ell_1 m_1 ; \ell_2 m_2} = (-1)^{m_{2}}\sqrt{\frac{(2\ell_1+1)(2\ell_2+1)}{4\pi}}\nonumber \\
  &\sum_{\ell_3=0}^{2\ell_{\mathrm{max}}}a_{\ell_3 m_3} \sqrt{2\ell_3+1}\,
  \left(\begin{array}{ccc}
      \ell_3 & \ell_1 & \ell_2 \\
      \\
      \,0\quad & \,\,0\quad & \,0\,
    \end{array}\right)
  \left(\begin{array}{ccc}
      \ell_3 & \ell_1 & \ell_2 \\
      \\
      m_3\quad    & m_1\,    & -m_2\,
    \end{array}\right)
  \nonumber \\
  &\,\delta(m_3 + m_1 - m_2)
\end{align}
almost identical with Eq. (\ref{eq:Nlmmat}). As in the previous case the internal sum covers multipoles up to $2\ell_{\mathrm{max}}$,
and the main difference is a change in sign of $m_1,m_2$.
We find that the mask coupling matrix is relatively well conditioned and further manipulations are not necessary. This has been tested by simply inverting the kernel 
matrix $\bfW$ constructed from the resulting mask which is used in the analysis of the WMAP data.

\subsection{Likelihood maximization scheme}
\label{sec:maxlikelihood}

In order to maximize the likelihood we use a non-linear Newton-Rapson search algorithm\footnote{using dmng.f from www.netlib.org} for the direction and amplitude. 
Finding the maximum of the likelihood 
is equivalent to finding the minimum of the quantity
\beq
\label{eq:logL}
-2\log{\mL} = \bfd^\dagger \bfC^{-1} \bfd + \mathrm{Tr} \log \bfC.
\eeq
which has a global minimum when the likelihood function has a global maximum.
The algorithm minimizes a general unconstrained function by evaluating the first and second derivatives. 
Due to the symmetry of the signal covariance matrix, $\varpi_0$ 
is constrained to be larger or equal to zero. A negative amplitude
can always be replaced with a positive one and a shift in the angles
$(\theta,\phi)$: $\mathbf{S}(-\varpi_0,n_{\mathrm{de}},\theta,\phi) = \mathbf{S}(\varpi_0,n_{\mathrm{de}},\pi-\theta,\pi+\phi)$. This is easily
seen from the definition of the signal covariance matrix.

The gradient is computed analytically, the derivative of Eq. (\ref{eq:logL}) with respect to any of the parameters in the set $\al$ is
\begin{widetext}
\begin{equation}
  \frac{\partial (-2\log{\mL})}{\partial\al} = \bfd^{\dagger} \frac{\partial \bfC^{-1}}{ \partial \al}\bfd
  + \mbox{Tr}\left(\frac{\partial \log \bfC}{\partial \al}\right) =
  -\bfd^{\dagger} \bfC^{-1} \frac{\partial \bfC}{ \partial \al}\,\bfC^{-1}\bfd
  + \mathrm{Tr}\left(\bfC^{-1}\frac{\partial \bfC}{\partial \al}\right)
  \label{eq:dlogL}
\end{equation}
\end{widetext}
To find the derivative of $\bfC^{-1}$ we have differentiated the identity matrix $\bf{I} = \bfC \bfC^{-1}$ and 
solved for the derivative of the inverse covariance matrix in terms of the derivative of the matrix itself.
The analytic derivative of $\bfC$ is computed from Eq. \ref{corr}. The second derivatives are
estimated from the gradient using a secant method. Depending on our initial guess for the parameters and
our convergence criteria the minimizer in general may or may not converge
to a global minimum. In our case false convergence is rarely a problem since the likelihood surface is well behaved and the 
local minimum has a small amplitude compared to the global.

For the spectral index $n_\mathrm{de}$ we run a grid calculation. In each grid point, we apply the above maximization procedure and find 
the value of the maximum likelihood for the given value of $n_\mathrm{de}$. In the end we search the grid to find the full global maximum in the 4-parameter space.

\section{Application to WMAP-data}
\label{sec:results}

Let us now discuss the results obtained with and without normalization of our signal covariance matrix (see section \ref{sect:modspec}).
We analyze the V-band (61 GHz) data map which is believed to be one of the cleanest bands in terms of foreground residuals, and
recommended for cosmological analysis by the WMAP team. To this map we apply the modified WMAP KQ$85$ galactic skycut, removing $25\%$ of the sky.
 Since we are only analyzing the largest scales no special care is taken with regards to point-source masking.
We take into account the noise RMS pattern and the corresponding beam properties for the V-band. We analyze these maps out to a
maximum multipole moment $\ell_{\mathrm{max}} = 20$. The kernel matrices $\bfW$ 
which emulate the effect of a skycut in harmonic space includes multipoles up to $\ell_{\mathrm{max}}=40$. Now that we have our 
data map we are ready to start the analysis.

\subsection{Unnormalized covariance matrix}
\label{sec:no_norm_c}

\begin{figure}[t]
  \mbox{\epsfig{figure=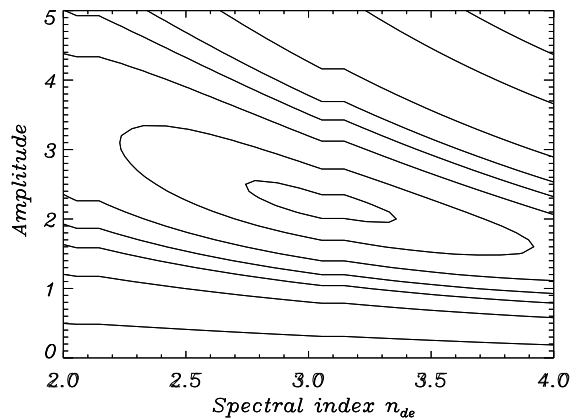,width=\linewidth,clip=}}
  \caption{2-dimensional plot of the raw likelihood (Posterior distribution)
    as a function of dark energy spectral index $n_{\mathrm{de}}$ and the transformed amplitude $\varpi^a_0$, using
    WMAP 7-yr data. The spherical angles have been fixed to their WMAP best-fit values at $\ell_{\mathrm{max}}=20$ to enable
    a projection. The posterior mode value is located at $n_{\mathrm{de}}=3.1$. The (transformed) amplitude found in this plot is higher 
    than the value for the true anisotropic amplitude noted in table \ref{tb:reswmap1} due to a bias introduced by application of the mask
    and a linear transformation.}
\label{fig:2dplot}
\end{figure}

Performing a grid-calculation with a spectral index range of $-5\leq n_{\mathrm{de}} \leq 5$ with a stepsize of $\Delta n_{\mathrm{de}}=0.1$, 
where for each value of $n_{\mathrm{de}}$ we do a 3-dimensional search for the peak of the likelihood function using our
likelihood maximization scheme described in section \ref{sec:maxlikelihood} we find that the likelihood
for negative spectral index values are very insignificant. As we approach $n_{\mathrm{de}} = 0$ the likelihood
starts peaking slowly until we find a peak at $n_{\mathrm{de}} = 3.1$. The best-fit direction
remains practically constant as me move through the grid (the change is completely negligible compared to the uncertainty) indicating that 
correlation between the two sets $(\theta,\phi)$ and $(n_{\mathrm{de}},\varpi_0)$ is weak.

We find Fisher matrix error bars calculating the Fisher matrix using
\beq
\label{eq:fishtr}
F_{\alpha \beta} = \frac{1}{2}\mathrm{Tr}\left(\frac{dC}{d\lambda_\alpha}C^{-1}\frac{dC}{d\lambda_\beta}C^{-1} \right)
\eeq
where the derivatives of the covariance matrix are found analytically for the direction and amplitude
and numerically for the spectral index. Since the amplitude and spectral index are weakly correlated,
the off-diagonal elements are taken into account in the matrix. The results are shown in table \ref{tb:reswmap1}.
As expected, in order for the model to be consistent with the power spectrum, we find $\varpi_0$ consistent with
zero within the 1$\sigma$ error. In figure \ref{fig:2dplot} we show the likelihood surface close to the peak. Note that the amplitude in this plot is the transformed amplitude  $\varpi^a_0$ (see section \ref{sect:trans}) and that the amplitude at the maximum of the likelihood is biased with respect to the best fit amplitude. This bias results from the complicated form of the likelihood introduced by the mask. The bias is corrected for in the following manner: given the parameters found from the peak of the likelihood, we generate 100 anisotropic realisations. For each realisation we estimate the anisotropic parameters and in the end
compute the average bias in $\varpi_0^a$. Next we subtract the bias from the input-value  and repeat the procedure until our average computed
amplitude matches the value found in the WMAP data. When the bias has been subtracted we are left with "the true"
estimate of the transformed amplitude $\varpi_0^a$. The fiducial amplitude is then obtained using $\varpi_0 = \sqrt{a}\varpi_0^a$. In the unnormalized
case we find a final amplitude value $\varpi_0 = 0.51$.

\begin{figure}[t]
  \mbox{\epsfig{figure=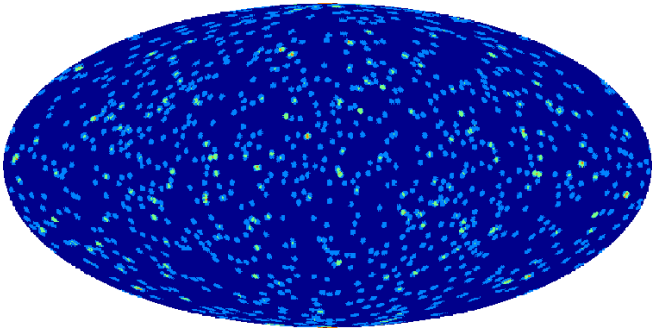, width=\linewidth,clip=}}
  \caption{Distribution of $(\theta,\phi)$ values on the sphere from Gaussian ($\varpi_0 = 0$ input)
    simulations. Note how the positions found in the simulations are randomly distributed on the sphere
  and not aligned along some particular axis, in clear agreement with a random Gaussian distribution.}
\label{fig:angdist}
\end{figure}

The best fit direction is somewhat close to the galactic center. In order to check that this is not caused by the shape of the mask, we estimated the direction on 1000 simulated isotropic maps and found that there is no bias towards the galactic center. The estimated directions are shown in figure \ref{fig:angdist}.

\begin{figure}[t]
  \mbox{\epsfig{figure=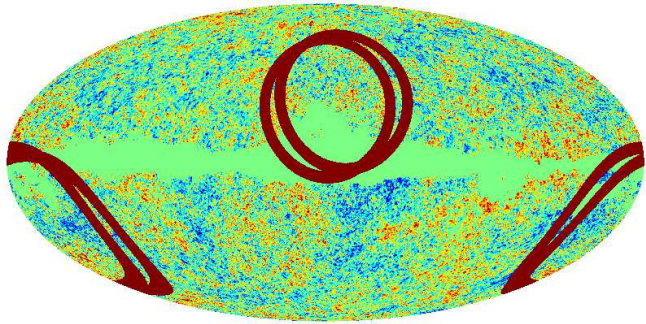, angle=90,width=\linewidth,clip=}}
  \caption{Map indicating the $1\sigma$ uncertainty in the preferred direction of the axis. The background is the V-band (61 GHz) WMAP
    7-yr data map with the KQ85 mask. The two axes plotted are the directions given in table \ref{tb:reswmap1} for the unnormalized and
    the normalized cases.}
\label{fig:prefdir}
\end{figure}

The error in the amplitude has also been estimated using $1000$ Monte Carlo simulations. We find
that the error from simulations agrees well with the error found using the Fisher matrix.
In figure \ref{fig:prefdir} we show the best fit direction with error bars.

\begin{table}[t]
  \caption{Results from WMAP 7yr-data. Lower line shows the results with normalized covariance matrix.}
  \centering
  \begin{tabular}{c|c|c}
    \hline
    $n_{\mathrm{de}}$ & $\varpi_0$ & $\hbn \,(l,b)$ \\
    \hline \hline
    $3.1 \pm 1.5$ & $0.51 \pm 0.94$  & $(168^\circ,-31^\circ)$ \\
    $1.2 \pm 0.7$ & $7.12 \pm 3.82 $  & $(179^\circ,-27^\circ)$ \\
  \end{tabular}
  \label{tb:reswmap1}
\end{table}

An isotropic universe is clearly preferred by the data using this model.

\subsection{Normalized covariance matrix}
\label{sec:norm_c}

The model with an unnormalized matrix (see Eq.\ref{corr}) giving the modified power spectrum in Eq. (\ref{clmod}) is clearly
not preferred by the WMAP 7-yr data. However, if we allow other cosmological parameters to vary we may be able to find
a better fit as explained above. We have therefore repeated the procedure using the normalization in Eq. (\ref{normSignal}) 
which means fixing the diagonal part of the covariance matrix to the best fit $C_\ell$ regardless of amplitude and spectral index.

The lower line in table \ref{tb:reswmap1} shows the results for the anisotropic cosmological parameters from the exploration of the 
likelihood space with a normalized covariance matrix. Again, a grid was set up for the spectral index in the interval
$-5\le n_{\mathrm{de}} \le 5$ with the same stepsize and for each value of $n_{\mathrm{de}}$ we estimated the best-fit values of $(\theta,\phi,\varpi_0)$ and
the corresponding value of the likelihood. The preferred amplitude in the normalized case is so large that a huge change 
of cosmological parameter values would be necessary to obtain the best fit WMAP spectrum. No physical solution is found in the normalized case.

\section{Conclusions}

In this work we tested anisotropic dark energy models with the 7-year WMAP temperature observations data. If dark energy is not a perfect fluid but for instance a vector field,
the CMB sky will be distorted anisotropically on its way to us by the ISW effect. The signal covariance matrix then becomes nondiagonal for
small multipoles, but at $\ell \gtrsim 20$ the anisotropy is negligible. This can be used to constrain violations of rotational invariance
in the late universe, and to obtain hints on possible imperfect nature of dark energy and the large-angle anomalous features in the CMB.

To model this phenomenon, we introduced a mismatch of the two gravitational potentials in the Newtonian gauge. The mismatch, quantified by $\varpi$, is proportional to a gradient
along the preferred axis $\hat{n}$. We also allowed this effect to depend on the scale by introducing the spectral index $n_{\mathrm{de}}$. Physically, such gradient could be caused by a
large-scale
inhomogeneity entering our horizon, spontaneous formation due to e.g. coherent magnetic fields or simply the possible imperfect nature of the dark energy field. Many possible
realisations of the latter possibility were discussed in the introduction and in the section \ref{sec:ade}. The dominant effect on the CMB is then a quadropole modulation, which has
the same geometrical correlation structure but different time and scale dependence than in models considered previously. Now a dipole modulation, though subdominant, is predicted
too.

We calculate the mode couplings introduced to the spherical harmonic coefficients of the CMB by the anisotropic model and obtain the full likelihood for the lowest multipoles where the dominant contribution to the model is expected to be found. Maximizing the likelihood taking into account the instrumental parameters of the WMAP experiment, we are able to find optimal estimates of the anisotropic parameters.
Analysis of the masked WMAP V-band, fixing other cosmological parameters, gave a best fit amplitude $\varpi_0=0.51\pm0.94$ and $n_{\mathrm{de}} = 3.1$ consistent with an isotropic universe. 

In comparison, test of the isotropic version of this parametrisation show that the data is then compatible with a vanishing deviation, and
allows a nonzero $\varpi$ of the order of $\mathcal{O}(0.1)$ (\cite{Daniel:2010ky}). At the level of Solar system, no hints of deviations are observed, and the post-Newtonian
correction is constrained to be at most $\mathcal{O}(10^{-5})$ (\cite{Will:2001mx}). However, the numbers themselves cannot be directly compared, since our best-fit model features also a strong
scale-dependence of the deviation. The preferred steep blue spectral index may be due to the fact that the statistically most significant contribution must come from
smaller scales, the largest scales of the CMB being severely cosmic variance limited.

Another shortcoming of our parameterization is its inability to incorporate the lack of large-angle 
power in the observed sky, one of the most striking anomaly present in the data.
Indeed, the main lesson to be derived from our study is that cosmological constraints on realistic 
imperfect post-Newtonian deviations are on the order $\mathcal{O}(10^{-4})$ (reported in table \ref{tb:reswmap1}), only an order 
of magnitude below those obtained from the Solar system scales. This motivates to further investigate the possible origin and constraints of 
imperfect source terms in cosmology. In particular, a fully consistent post-Friedmannian parametrisation along the lines of \cite{ferreira,Baker:2011jy}, 
tailored to the study of directional dependence of deviations from the standard predictions of linearised cosmology, remains to be developed.

\subsection{acknowledgements}
We thank Hans Kristian Eriksen for useful discussions. The work of TK was
supported by the Academy of Finland and the Yggdrasil grant from the
Norwegian Research Council. DFM and FKH thank the Research Council of Norway for FRINAT grant 197251/V30 and an OYI-grant respectively. DFM is also partially supported by project  PTDC/FIS/111725/2009 and CERN/FP/123618/2011.
Maps and results have been derived using the Healpix\footnote{http://healpix.jpl.nasa.gov} software package developed by \cite{Gorski:2004by}.
The anisotropic transfer functions have been derived using a modified version of CAMB due to \cite{Lewis:1999bs}.
We acknowledge the use of the LAMBDA archive (Legacy Archive for Microwave Background Data Analysis). Support for LAMBDA is provided by the NASA office for Space Science. 

\clearpage

\appendix

\section{Anisotropic contribution to the power spectrum}
The diagonal part of the covariance matrix is a sum of the power spectrum due to the isotropy and 
a term determined by the anisotropic parameters $\varpi_0$, $n_{\mathrm{de}}$ and $(\theta,\phi)$:
\beq
S_{\ell m; \ell m} = \frac{2}{\pi} \left( I_{\ell} + \varpi^2_0 \xi_{\ell m;\ell m} I^{\mathrm{AA}}_{\ell \ell} \right)
= C_\ell + \xi_{\ell m;\ell m} \varpi^2_0 C^{\mathrm{AA}}_{\ell } 
\eeq
The dependence on the spectral index comes from the integral over the anisotropic transfer functions in $I^{\mathrm{AA}}_{\ell \ell}$.
The diagonal part of the geometric factor is \citep{Ackerman:2007nb}
\beq
\xi_{\ell m;\ell m} = -2n_{+}n_{-}\frac{-1+\ell(\ell+1)+m^2}{(2\ell -1)(2\ell + 3)} + n^2_{0} \frac{2\ell(\ell +1)-2m^2 -1}{(2\ell -1)(2\ell + 3)}
\eeq
where the spherical components $n_{+},n_{-},n_0$ containing the angular dependence have been defined in Equation (\ref{vector}). 
Using the well-known result
\beq
 \sum_{m=1}^{\ell} m^2 = \frac{\ell(\ell +1)(2\ell+1)}{6}
\eeq
we find that the average of the geometric factor is free from angular dependence and simplifies nicely to
\beq
\sum_{m=-\ell}^{\ell} \xi_{\ell m;\ell m} = \frac{2\ell + 1}{3}
\eeq
With this result one finds that the theoretical prediction for the modified power spectrum due to the anisotropic component
becomes
\beq
C^{\mathrm{mod}}_\ell = \frac{1}{2\ell+1}\sum_{m} \langle a_{\ell m} a^*_{\ell m} \rangle = C_{\ell} + \frac{\varpi^2_0}{3}\,C^{\mathrm{AA}}_{\ell}
\eeq
which is the modified power spectrum $C^{\mathrm{mod}}_\ell$ quoted in Eq. (\ref{clmod}).

\bibliography{arefs}

\begin{thebibliography}{70}
\expandafter\ifx\csname natexlab\endcsname\relax\def\natexlab#1{#1}\fi

\bibitem[{Abramo \& Pereira(2010)}]{Abramo:2010gk}
Abramo, L.~R., \& Pereira, T.~S. 2010, arXiv: 1002.3173 [astro-ph.CO]

\bibitem[{Ackerman {et~al.}(2007)Ackerman, Carroll, \& Wise}]{Ackerman:2007nb}
Ackerman, L., Carroll, S.~M., \& Wise, M.~B. 2007, Phys. Rev., D75, 083502

\bibitem[{Adamek {et~al.}(2010)Adamek, Campo, \& Niemeyer}]{Adamek:2010sg}
Adamek, J., Campo, D., \& Niemeyer, J.~C. 2010, arXiv: 1003.3204 [hep-th]

\bibitem[{Afshordi {et~al.}(2009)Afshordi, Geshnizjani, \&
  Khoury}]{Afshordi:2008rd}
Afshordi, N., Geshnizjani, G., \& Khoury, J. 2009, JCAP, 0908, 030

\bibitem[{Akarsu \& Kilinc(2010)}]{ani5}
Akarsu, O., \& Kilinc, C.~B. 2010, Gen. Rel. Grav., 42, 763

\bibitem[{Aluri \& Jain(2011)}]{Aluri:2011wv}
Aluri, P.~K., \& Jain, P. 2011, arXiv: 1108.5894 [astro-ph.CO]

\bibitem[{Appleby {et~al.}(2010)Appleby, Battye, \& Moss}]{bat1}
Appleby, S., Battye, R., \& Moss, A. 2010, Phys. Rev., D81, 081301

\bibitem[{Armendariz-Picon(2004)}]{ArmendarizPicon:2004pm}
Armendariz-Picon, C. 2004, JCAP, 0407, 007

\bibitem[{Armendariz-Picon \& Pekowsky(2008)}]{ArmendarizPicon:2008yr}
Armendariz-Picon, C., \& Pekowsky, L. 2008, 0807.2687

\bibitem[{Baker {et~al.}(2011)Baker, Ferreira, Skordis, \&
  Zuntz}]{Baker:2011jy}
Baker, T., Ferreira, P.~G., Skordis, C., \& Zuntz, J. 2011, arXiv: 1107.0491
  [astro-ph.CO]

\bibitem[{Battye \& Moss(2009)}]{bat2}
Battye, R., \& Moss, A. 2009, Phys. Rev., D80, 023531

\bibitem[{Bennett {et~al.}(2003)}]{Bennett:2003bz}
Bennett, C.~L., {et~al.} 2003, Astrophys. J. Suppl., 148, 1

\bibitem[{Boehmer \& Mota(2008)}]{boehmer}
Boehmer, C.~G., \& Mota, D.~F. 2008, Phys. Lett., B663, 168

\bibitem[{Caldwell {et~al.}(2007)Caldwell, Cooray, \&
  Melchiorri}]{Caldwell:2007cw}
Caldwell, R., Cooray, A., \& Melchiorri, A. 2007, astro-ph/0703375

\bibitem[{Campanelli(2009)}]{ani2}
Campanelli, L. 2009, Phys. Rev., D80, 063006

\bibitem[{Cooke \& Lynden-Bell(2009)}]{ani6}
Cooke, R., \& Lynden-Bell, D. 2009, arXiv: 0909.3861[astro-ph.CO]

\bibitem[{Cooray {et~al.}(2008)Cooray, Holz, \& Caldwell}]{ani4}
Cooray, A.~R., Holz, D.~E., \& Caldwell, R. 2008, astro-ph/0812.0376

\bibitem[{Copi {et~al.}(2010)Copi, Huterer, Schwarz, \& Starkman}]{Copi:2010na}
Copi, C.~J., Huterer, D., Schwarz, D.~J., \& Starkman, G.~D. 2010, arxiv:
  1004.5602[astro-ph.CO]

\bibitem[{Daniel {et~al.}(2008)Daniel, Caldwell, Cooray, \&
  Melchiorri}]{Daniel:2008et}
Daniel, S.~F., Caldwell, R.~R., Cooray, A., \& Melchiorri, A. 2008, Phys. Rev.,
  D77, 103513

\bibitem[{Daniel {et~al.}(2009)Daniel, Caldwell, Cooray, Serra, \&
  Melchiorri}]{Daniel:2009kr}
Daniel, S.~F., Caldwell, R.~R., Cooray, A., Serra, P., \& Melchiorri, A. 2009,
  0901.0919

\bibitem[{Daniel {et~al.}(2010)}]{Daniel:2010ky}
Daniel, S.~F., {et~al.} 2010, arxiv: 1002.1962 [astro-ph.CO]

\bibitem[{Dimastrogiovanni {et~al.}(2008)Dimastrogiovanni, Fischler, \&
  Paban}]{ani3}
Dimastrogiovanni, E., Fischler, W., \& Paban, S. 2008, JHEP, 07, 045

\bibitem[{Dimopoulos {et~al.}(2010)Dimopoulos, Karciauskas, \&
  Wagstaff}]{Dimopoulos:2009am}
Dimopoulos, K., Karciauskas, M., \& Wagstaff, J.~M. 2010, Phys.Rev., D81,
  023522

\bibitem[{Dimopoulos {et~al.}(2011)Dimopoulos, Wills, \&
  Zavala}]{Dimopoulos:2011pe}
Dimopoulos, K., Wills, D., \& Zavala, I. 2011, arXiv: 1108.4424 [hep-th]

\bibitem[{Dvali {et~al.}(2000)Dvali, Gabadadze, \& Porrati}]{dvali}
Dvali, G.~R., Gabadadze, G., \& Porrati, M. 2000, Phys. Lett., B485, 208

\bibitem[{Erickcek {et~al.}(2008)Erickcek, Kamionkowski, \&
  Carroll}]{Erickcek:2008sm}
Erickcek, A.~L., Kamionkowski, M., \& Carroll, S.~M. 2008, Phys. Rev., D78,
  123520

\bibitem[{Eriksen {et~al.}(2004)Eriksen, Hansen, Banday, Gorski, \&
  Lilje}]{Eriksen:2003db}
Eriksen, H.~K., Hansen, F.~K., Banday, A.~J., Gorski, K.~M., \& Lilje, P.~B.
  2004, Astrophys. J., 605, 14

\bibitem[{Ferreira \& Skordis(2010)}]{ferreira}
Ferreira, P.~G., \& Skordis, C. 2010, arxiv: 1003.4231 [astro-ph.CO]

\bibitem[{Germani \& Kehagias(2009)}]{Germani:2009iq}
Germani, C., \& Kehagias, A. 2009, JCAP, 0903, 028

\bibitem[{Golovnev {et~al.}(2008)Golovnev, Mukhanov, \&
  Vanchurin}]{Golovnev:2008cf}
Golovnev, A., Mukhanov, V., \& Vanchurin, V. 2008, JCAP, 0806, 009

\bibitem[{Gordon {et~al.}(2005)Gordon, Hu, Huterer, \&
  Crawford}]{Gordon:2005ai}
Gordon, C., Hu, W., Huterer, D., \& Crawford, T.~M. 2005, Phys. Rev., D72,
  103002

\bibitem[{Gorski {et~al.}(2005)}]{Gorski:2004by}
Gorski, K.~M., {et~al.} 2005, Astrophys. J., 622, 759

\bibitem[{Graham {et~al.}(2010)Graham, Harnik, \& Rajendran}]{Graham:2010hh}
Graham, P.~W., Harnik, R., \& Rajendran, S. 2010, arxiv: 1003.0236 [hep-th]

\bibitem[{Groeneboom {et~al.}(2010)Groeneboom, Axelsson, Mota, \&
  Koivisto}]{Groeneboom:2010fn}
Groeneboom, N., Axelsson, M., Mota, D., \& Koivisto, T. 2010, arXiv: 1011.5353
  [astro-ph.CO]

\bibitem[{{Groeneboom} {et~al.}(2010){Groeneboom}, {Axelsson}, {Mota}, \&
  {Koivisto}}]{Groeneboom:2010hp}
{Groeneboom}, N.~E., {Axelsson}, M., {Mota}, D.~F., \& {Koivisto}, T. 2010,
  ArXiv e-prints

\bibitem[{Groeneboom \& Eriksen(2008)}]{Groeneboom:2008fz}
Groeneboom, N.~E., \& Eriksen, H.~K. 2008, 0807.2242

\bibitem[{Gumrukcuoglu {et~al.}(2006)Gumrukcuoglu, Contaldi, \&
  Peloso}]{Gumrukcuoglu:2006xj}
Gumrukcuoglu, A.~E., Contaldi, C.~R., \& Peloso, M. 2006, astro-ph/0608405

\bibitem[{Gumrukcuoglu {et~al.}(2007)Gumrukcuoglu, Contaldi, \&
  Peloso}]{Gumrukcuoglu:2007bx}
---. 2007, JCAP, 0711, 005

\bibitem[{Hansen {et~al.}(2008)Hansen, Banday, Gorski, Eriksen, \&
  Lilje}]{Hansen:2008ym}
Hansen, F.~K., Banday, A.~J., Gorski, K.~M., Eriksen, H.~K., \& Lilje, P.~B.
  2008, 0812.3795

\bibitem[{Hinshaw {et~al.}(2007)}]{Hinshaw:2006ia}
Hinshaw, G., {et~al.} 2007, Astrophys. J. Suppl., 170, 288

\bibitem[{Hoftuft {et~al.}(2009)}]{Hoftuft:2009rq}
Hoftuft, J., {et~al.} 2009, 0903.1229

\bibitem[{Hu \& Sawicki(2007)}]{hu}
Hu, W., \& Sawicki, I. 2007, Phys. Rev., D76, 104043

\bibitem[{Jimenez {et~al.}(2009)Jimenez, Koivisto, Maroto, \& Mota}]{vec10}
Jimenez, J.~B., Koivisto, T.~S., Maroto, A.~L., \& Mota, D.~F. 2009, JCAP,
  0910, 029

\bibitem[{Jimenez \& Maroto(2008)}]{Jimenez:2008au}
Jimenez, J.~B., \& Maroto, A.~L. 2008, Phys. Rev., D78, 063005

\bibitem[{Jimenez \& Maroto(2009{\natexlab{a}})}]{Jimenez:2008nm}
---. 2009{\natexlab{a}}, JCAP, 0903, 016

\bibitem[{Jimenez \& Maroto(2009{\natexlab{b}})}]{Jimenez:2009dt}
---. 2009{\natexlab{b}}, 0903.4672

\bibitem[{Kiselev(2004)}]{Kiselev:2004py}
Kiselev, V.~V. 2004, Class. Quant. Grav., 21, 3323

\bibitem[{Koivisto \& Mota(2006)}]{Koivisto:2005mm}
Koivisto, T., \& Mota, D.~F. 2006, Phys. Rev., D73, 083502

\bibitem[{Koivisto \& Mota(2008{\natexlab{a}})}]{Koivisto:2008ig}
---. 2008{\natexlab{a}}, JCAP, 0806, 018

\bibitem[{Koivisto \& Mota(2008{\natexlab{b}})}]{Koivisto:2008xf}
Koivisto, T.~S., \& Mota, D.~F. 2008{\natexlab{b}}, JCAP, 0808, 021

\bibitem[{Koivisto \& Mota(2011)}]{Koivisto:2010fk}
---. 2011, JHEP, 1102, 061

\bibitem[{Koivisto {et~al.}(2009)Koivisto, Mota, \& Pitrou}]{Koivisto:2009sd}
Koivisto, T.~S., Mota, D.~F., \& Pitrou, C. 2009, 0903.4158

\bibitem[{Koivisto {et~al.}(2011)Koivisto, Mota, Quartin, \& Zlosnik}]{zspace}
Koivisto, T.~S., Mota, D.~F., Quartin, M., \& Zlosnik, T.~G. 2011, Phys.Rev.,
  D83, 023509

\bibitem[{Koivisto \& Nunes(2009)}]{Koivisto:2009fb}
Koivisto, T.~S., \& Nunes, N.~J. 2009, Phys. Rev., D80, 103509

\bibitem[{Land \& Magueijo(2005)}]{Land:2005ad}
Land, K., \& Magueijo, J. 2005, Phys. Rev. Lett., 95, 071301

\bibitem[{Lewis {et~al.}(2000)Lewis, Challinor, \& Lasenby}]{Lewis:1999bs}
Lewis, A., Challinor, A., \& Lasenby, A. 2000, Astrophys. J., 538, 473

\bibitem[{Li {et~al.}(2008)Li, Fonseca~Mota, \& Barrow}]{Li}
Li, B., Fonseca~Mota, D., \& Barrow, J.~D. 2008, Phys. Rev., D77, 024032

\bibitem[{Libanov {et~al.}(2007)Libanov, Rubakov, Papantonopoulos, Sami, \&
  Tsujikawa}]{Libanov:2007mq}
Libanov, M., Rubakov, V., Papantonopoulos, E., Sami, M., \& Tsujikawa, S. 2007,
  JCAP, 0708, 010

\bibitem[{Manera \& Mota(2006)}]{manera}
Manera, M., \& Mota, D. 2006, Mon.Not.Roy.Astron.Soc., 371, 1373

\bibitem[{Mota {et~al.}(2007)Mota, Kristiansen, Koivisto, \&
  Groeneboom}]{Mota:2007sz}
Mota, D.~F., Kristiansen, J.~R., Koivisto, T., \& Groeneboom, N.~E. 2007, Mon.
  Not. Roy. Astron. Soc., 382, 793

\bibitem[{Pereira {et~al.}(2007)Pereira, Pitrou, \& Uzan}]{Pereira:2007yy}
Pereira, T.~S., Pitrou, C., \& Uzan, J.-P. 2007, JCAP, 0709, 006

\bibitem[{Prunet {et~al.}(2005)Prunet, Uzan, Bernardeau, \&
  Brunier}]{Prunet:2004zy}
Prunet, S., Uzan, J.-P., Bernardeau, F., \& Brunier, T. 2005, Phys. Rev., D71,
  083508

\bibitem[{Rakic \& Schwarz(2007)}]{Rakic:2007ve}
Rakic, A., \& Schwarz, D.~J. 2007, Phys. Rev., D75, 103002

\bibitem[{Rodrigues(2008)}]{ani1}
Rodrigues, D.~C. 2008, Phys. Rev., D77, 023534

\bibitem[{Rubakov(2006)}]{Rubakov:2006pn}
Rubakov, V.~A. 2006, Theor. Math. Phys., 149, 1651

\bibitem[{Skordis(2009)}]{skordis}
Skordis, C. 2009, Phys. Rev., D79, 123527

\bibitem[{Tangen(2009)}]{Tangen:2009mw}
Tangen, K. 2009, arxiv: 0910.4164 [astro-ph.CO]

\bibitem[{Will(2001)}]{Will:2001mx}
Will, C.~M. 2001, Living Rev. Rel., 4, 4

\bibitem[{Zlosnik(2011)}]{Zlosnik:2011iu}
Zlosnik, T. 2011, arXiv: 1107.0389 [gr-qc]

\bibitem[{Zuntz {et~al.}(2010)Zuntz, Zlosnik, Bourliot, Ferreira, \&
  Starkman}]{vec1}
Zuntz, J., Zlosnik, T.~G., Bourliot, F., Ferreira, P.~G., \& Starkman, G.~D.
  2010, arxiv: 1002.0849 [astro-ph.CO]

\end{thebibliography}

\end{document}